# Assessment of an analytical density-matrix derived from a modified Colle-Salvetti approach to the electron gas

Sébastien RAGOT

IBM Research – Zurich, Säumerstr. 4, 8803 Rüschlikon, Switzerland

## Abstract

The Ragot–Cortona model of local correlation energy (J. Chem. Phys. **121**, 7671 (2004)) revisits the initial approach of Colle and Salvetti [Theo. Chim. Acta **37**, 329 (1975)] in order to reinstate the kinetic contribution $T_c$ to the total correlation energy $E_c$. In this work, the one-electron reduced density-matrix underlying the amended model is fully derived in closed-form. By construction, the said density-matrix is parameter-free but not *N*-representable, owing to a series of approximations used in the Ragot–Cortona approach. However, the resulting density-matrix is shown to have formally correct short- and long-range expansions. Furthermore, its momentum-space counterpart qualitatively agrees with known parameterized momentum distributions except at small momenta, where the disagreement reflects the non-representability of the model and restricts to a small fraction of the slowest electrons only.

**Keywords**: Density-matrix, momentum distribution, uniform electron gas.

# I. Motivations

Systems of uniform charge density form an important class for the validation and the development of electronic structure theories. For example, the Density Functional Theory (DFT) is well known to reformulate the quantum mechanics of many-body systems in terms of functionals of the charge density $\rho$ rather than a many-body wave function [1, 2]. Since $\rho$ is thereby instated as a basic variable, systems of uniform charge density serve as a convenient basis for designing the said functionals, which are usually unknown.

A famous example is the quantal exchange energy of uniform non-interacting fermions, which find a very simple relationship with $\rho$. Likewise, the uniform electron gas (UEG) provides an ideal prototype of interacting electrons. How electronic properties of the UEG relate to $\rho$ can further be quite accurately exported to non-uniform systems, as in the local density approximation (LDA) or, more generally, its extension to spin-polarized states (LSDA).

Amongst the most successful post-local approximations, the generalized gradient approximations (GGAs) and its descendants drove DFT to a level of accuracy suited for computational chemistry and, understandably, attracted lots of investigations. Meanwhile, refining the standard LDA has remained a constant concern, giving rise to innumerable variations thereof. Examples are for instance schemes that remove spurious correlation energy contributions, e.g. the "self-correlation" contributions [3, 4] or approaches re-parameterizing the standard LDAs, see e.g. [5, 6].

Amongst other attempts, a modified local correlation energy functional was quite recently proposed, which was obtained from an approximate expression for the one-electron reduced density-matrix (or "density-matrix") of the interacting UEG [7]. In short, the approximate density-matrix was derived from an explicitly correlated wave function, as in the early model of Colle and Salvetti [8]. However, at variance with approaches based on pair densities, the total correlation energy was entirely obtained from the one-electron density-matrix, namely, from the correlation kinetic energy, by inverting the virial theorem. Because only the kinetic energy, *i.e.* the second momentum moment is needed, only a second-order series expansion of the density-matrix is required, whereby its entire analytical calculation can be skipped. The idea here was to remedy some of the drawbacks of the Colle–Salvetti model. In particular, correlation effects on the (properly normalized) density-matrix are reinstated. The final result is characterized by a simple analytical expression and does not contain any adjusted parameter.

The resulting correlation energy components were shown to be in satisfactory agreement with those of Perdew and Wang for the UEG [9]. As they further have remarkably simple analytical expressions, they can easily be implemented and tested in any DFT code. Calculations were first performed for atoms and ions [7]. Results for correlation energies and for ionization potentials were shown to substantially improve over the standard LDA.

More recently, Tognetti, Cortona, and Adamo [10] have closely investigated this local functional, which they refer to as "RC". First, its use in a purely local scheme ("RC04") has made it possible to reach a striking improvement of thermodynamic data of a large set of molecules (including heights of barriers for some chemical reactions), with respect to the standard local spin-density (LSDA) approximation. The authors conclude that the RC04 functional significantly reduces the gap between local and gradient-corrected approaches, if not cancels it.

This conclusion was confirmed in subsequent investigations [11 - 12], e.g. directed to ionization potentials of third row transition metal atoms and selected molecules, and extending the initial RC04 model to a gradient-corrected version ("TCA") [13], using a very simple ansatz for the gradient-correction enhancement factor. The TCA correlation functional was coupled with the Perdew–Burke–Erzernhof (PBE) exchange [14], which altogether was shown to significantly improve on the original PBE, at least for some physicochemical properties. Practically, Tognetti and coworkers have included the RC04 functional and descendants in a development version of the GAUSSIAN program package [15].

Quantitatively speaking, the mean absolute errors obtained with the purely local RC04 functional for the atomization energies of the G2-1 (55 molecules) and G2 (148 molecules) sets are of 14.6 and 26.3 kcal/mol, respectively, which can be compared with the LDA results (39.7 and 91.4), PBE (8.1 and 17.0), and TCA (6.7 and 9.0).

Such results are attractive. Explications inferred with respect to the original RC model is that it benefits from a sound physical background and/or fortuitous cancellations [10 - 13]. In this respect, a number of reservations can be emphasized regarding the underlying density-matrix model [7], namely:

- First, the density-matrix is approximated as a truncated expansion of the formally exact density-matrix and is therefore not $N$-representable. Most importantly, while two-electron interactions are exhaustively reflected in the model, $N$-electron effects ($N > 2$) are discarded; and

- second, the Hartree–Fock matrix (*i.e.* a spherical Bessel function) involved in the calculation was substituted with a gaussian ansatz, for simplicity. The correlation factor itself has a gaussian tail.

Thus, there are enough uncertainties in the RC model to contemplate, first, a full analytical derivation and, second, a proper assessment of the underlying density-matrix, which has not been carried out so far.

To this aim, the basic ideas of the RC model are briefly recalled in section 2. In section 3, an exact analytical expression is derived for the underlying density-matrix. Its limits are assessed with respect to theoretically exact counterparts in section 4, and the density-matrix obtained and its corresponding momentum distribution are compared with known models, as obtained from parameterized momentum distributions. Atomic units are used throughout and the same notations as in Ref. [7] are used.

## II. The Ragot–Cortona model of local correlation

First, a synthetical view of the RC approach should be given, be it for the clarity of the present purpose. The basic quantities needed are:

(i) The density-matrix $\rho_1(\mathbf{r};\mathbf{r}')$, which can also be expressed in terms of baricentric ($\mathbf{R} = (\mathbf{r} + \mathbf{r}')/2$) and relative ($\mathbf{s} = \mathbf{r} - \mathbf{r}'$) coordinates, *i.e.*,

$$\rho_1(\mathbf{r};\mathbf{r}') \equiv \rho_1(\mathbf{R},\mathbf{s}) ; \tag{1}$$

(ii) The reciprocal form factor (or "auto correlation function") [16], closely related to the density-matrix, which can be defined as

$$B(\mathbf{s}) = \int \rho_1(\mathbf{R},\mathbf{s}) d\mathbf{R} ; \tag{2}$$

such that

(iii) the momentum distribution

$$n(\mathbf{p}) = \frac{1}{(2\pi)^3} \int \rho_1(\mathbf{R},\mathbf{s}) e^{i\mathbf{p}\cdot\mathbf{s}} d\mathbf{R} d\mathbf{s}, \tag{3}$$

is simply the 3D-Fourier transform of $B(\mathbf{s})$, and conversely:

$$B(\mathbf{s}) = \int n(\mathbf{p}) e^{-i\mathbf{p}\cdot\mathbf{s}} d\mathbf{p}. \tag{4}$$

The derivation of the RC functional relies first and foremost on a truncated expansion of the correlated density-matrix. More exactly, starting from a pairwise correlated wave function ansatz, i.e.,

$$\psi(x_1,...,x_N) = \psi^{HF}(x_1,...,x_N)\prod_{i<j}[1 - f(\mathbf{r}_i,\mathbf{r}_j)],$$

the density-matrix can be shown to develop as

$$\rho_1(\mathbf{r}_1;\mathbf{r}_1') = \eta\{\rho_{1,HF}(\mathbf{r}_1;\mathbf{r}_1') \quad (5)$$
$$+ 2\int[-f(\mathbf{r}_1,\mathbf{r}_2) - f(\mathbf{r}_1',\mathbf{r}_2) + f(\mathbf{r}_1,\mathbf{r}_2)f(\mathbf{r}_1',\mathbf{r}_2)]\rho_{2,HF}(\mathbf{r}_1,\mathbf{r}_2;\mathbf{r}_1',\mathbf{r}_2)d\mathbf{r}_2 + ...\},$$

where $\eta$ is a normalization factor, HF stands for Hartree–Fock approximation, $f$ is a two-particle correlation factor, and $\rho_{2,HF}(\mathbf{r}_1,\mathbf{r}_2;\mathbf{r}_1',\mathbf{r}_2)$ is a semi-diagonal two-electron reduced density-matrix at HF level.

Thus, the correlated density-matrix, Eq. (5), can be written as the HF density-matrix augmented by a correction which includes direct correlation effects of $N - 1$ electrons with a reference electron. Remarkably, the said correlation effects are accounted for *all together* in Eq. (5), at variance with the independent pair approximations inherent to most approaches based on the pair density. Practically, this means that Eq. (5) cannot be obtained from the correlated two-electron density-matrix model of Colle–Salvetti (Eq. (6) in their original paper). Subsequent terms in the expansion of the density-matrix involve increasing numbers of electrons in a same region of space and were not retained in the RC model.

Now, just as in the original model, a usual two-particle correlation function was used:

$$f(\mathbf{r}_1,\mathbf{r}_2) = \left\{1 - \Phi(\mathbf{R}_{12})\left(1 + \frac{r_{12}}{2}\right)\right\}e^{-\beta_c^2(\mathbf{R}_{12})r_{12}^2}, \quad (6)$$

wherein $\mathbf{R}_{12} = (\mathbf{r}_1 + \mathbf{r}_2)/2$, $\mathbf{r}_{12} = \mathbf{r}_2 - \mathbf{r}_1$, and $\beta_c^{-1}$ is a correlation radius, assumed to be proportional to the Seitz radius $r_s$, that is, $\beta_c(\mathbf{R}_{12}) = q\rho(\mathbf{R}_{12})^{1/3}$, where $q$ is the only parameter of the model. Likewise, $\Phi$ is to be determined: choosing for instance $\Phi = \sqrt{\pi}\beta_c/(1+\sqrt{\pi}\beta_c)$ amounts to canceling the first-order Coulomb correlation effects on the density-matrix [7].

Clearly, for a uniform system, each term under the integral sign in Eq. (5) involves relative coordinates only, notably $r_{12} = |\mathbf{r}_2 - \mathbf{r}_1|$ and $r_{1'2} = |\mathbf{r}_2 - \mathbf{r}_{1'}|$. Similarly, the density-matrix depends on $s = |\mathbf{r}_1 - \mathbf{r}_{1'}|$ only, and not on $\mathbf{R} = (\mathbf{r} + \mathbf{r'})/2$. Thus, the following notations can be adopted: $\rho_1(\mathbf{r}_1;\mathbf{r}_1') \equiv \rho_1(s)$ and $\rho_{2,HF}(\mathbf{r}_1,\mathbf{r}_2;\mathbf{r}_1',\mathbf{r}_2) \equiv \rho_{2,HF}(s,r_{12},r_{1'2})$.

In addition, it follows from Eq. (2) that $\rho_1(s)$ is proportional to the reciprocal form factor $B(s)$ in the case of a uniform system. The latter is thus conveniently normalized to unity, i.e. $B(0)=1$, which amounts to have the density-matrix normalized to the number of particles. However, the density-matrix of a uniform system is preferably normalized to unity in the litterature. Accordingly, the present definition of the reciprocal form factor $B(s)$ matches the usual definition of the density-matrix for uniform systems. Therefore, $B(s)$ shall be referred to as the "density-matrix" in the following.

The truncated density-matrix of Eq. (5) rewrites as

$$\rho_1(s) = \rho B(s) = \eta\rho\left[B_{HF}(s) + \frac{2}{\rho}\int[-f(r_{12}) - f(r_{1'2}) + f(r_{12})f(r_{1'2})]\rho_{2,HF}(s,r_{12},r_{1'2})d\mathbf{r}_2\right], \quad (7)$$

wherein

$$f(r_{12}) = \left\{1 - \Phi\left(1 + \frac{r_{12}}{2}\right)\right\}e^{-\beta_c^2 r_{12}^2}, \quad (8)$$

and $\beta_c$ and $\Phi$ depend on $\rho$ or, equivalently, on $r_s$ or $k_F$, that is, $\rho = \dfrac{3}{4\pi r_s^3} = \dfrac{k_F^3}{3\pi^2}$.

Then, the HF description of the UEG is known to result in

$$B_{HF}(s) = 3\frac{\sin(k_F s) - k_F s \cos(k_F s)}{(k_F s)^3} \qquad (9)$$

and

$$\rho_{2,HF}(s, r_{12}, r_{1'2}) = \frac{1}{2}\rho^2 \left[ B_{HF}(s) - \frac{1}{2} B_{HF}(r_{12}) B_{HF}(r_{1'2}) \right].$$

The function $B_{HF}(s)$ has a maximum $B_{HF}(0) = 1$ and oscillates to zero as $s \to \infty$. Thus, retaining only the short range behavior of $B_{HF}(s)$, i.e. $B_{HF}(s) \approx 1 - \beta_x^2 s^2 \approx e^{-\beta_x^2 s^2}$ ($\beta_x = C_x \rho^{1/3}$ and $C_x = 10^{-1/2}(3\pi^2)^{1/3}$, see Ref. [7]), $\rho_{2,HF}(s, r_{12}, r_{1'2})$ can be approximated as

$$\rho_{2,HF}(s, r_{12}, r_{1'2}) \approx \frac{1}{2}\rho^2 \left[ e^{-\beta_x^2 s^2} - \frac{1}{2} e^{-\beta_x^2 \left(2\sigma^2 + \frac{1}{2}s^2\right)} \right], \qquad (10)$$

where $\boldsymbol{\sigma} = \mathbf{r}_2 - \mathbf{R}$. As parameterized above, the gaussian approximation leads to an error of about 5% in the normalization condition of the HF pair distribution. Yet, it seems convenient inasmuch as correlation effects are expected to be most important near $r_{12} = 0$, at least when using a correlation factor as in Eq. (8).

Now, inserting Eq. (10) into Eq. (7) allows the derivation of analytic closed-form expressions, notably for the correlation kinetic energy. This can, for instance, be achieved by expanding first $B(s)$ up to second order and then making use of the relation $T = -\dfrac{3}{2} B''(0)$, whereby the calculation of the full density-matrix is avoided in that case.

Instead of being fitted, the only parameter involved, which determines the correlation-hole, was fixed using two limit conditions, (i) ensuring the correct sign for the kinetic energy $t_c$ when $r_s \to 0$, and (ii) maximizing $t_c$ at large $r_s$, leading to $q = 3(3\pi^2)^{1/3}/\sqrt{35}$. Once these conditions are satisfied, the model that results is parameter-free, in the same extent as the usual LDA is.

The resulting expression of the correlation kinetic energy $t_{c,RC}$ is, in numerical form:

$$t_{c,RC}(r_s) = \frac{1}{11.947492 + 14.906265 r_s + 4.844019 r_s^2}. \qquad (11)$$

The last step simply consists of inverting the virial theorem, i.e. $t_c = -\dfrac{\partial (r_s \varepsilon_c)}{\partial r_s}$. The correlation energy $\varepsilon_{c,RC}$ obtained is

$$\varepsilon_{c,RC} = \frac{-0.655868 \tan^{-1}(4.888270 + 3.177037 r_s) + 0.897889}{r_s}. \qquad (12)$$

Thus, the correlation functional has a very simple expression, and can easily be implemented in any DFT code.

For instance, it allows for recovering 84% of the correlation at energy at $r_s = 0.5$, a value that drops to under 72% for $r_s > 5$, see Fig. 3 of Ref. [7]. Interestingly, it improves on the correlation energy obtained from the Colle–Salvetti formulation of the correlated pair density, independently of the parameterization used. This has been attributed to a better description of the correlation, beyond the independent pair approximation.

# III. Analytical density-matrices for the electron gas

## III.1. Closed-form expression of the RC density-matrix

The approximate density-matrix of Eq. (7) rewrites as

$$\rho_1(s) = \eta\{\rho_{1,HF}(s) + \rho_{1,Corr}(s)\}, \tag{13}$$

where $\rho_{1,Corr}(s)$ is a correlation correction to be determined analytically. The above equation can be more conveniently restated, by virtue of $\rho_1(s) = \rho B(s)$, as

$$B(s) = \eta\{B_{HF}(s) + B_{Corr}(s)\}, \tag{14}$$

where $\eta$ is obtained by imposing the correct limit $B(0) = 1$. Hence, what remains to be determined is

$$B_{Corr}(s) = \frac{2}{\rho}\int [-f(r_{12}) - f(r_{1'2}) + f(r_{12})f(r_{1'2})]\rho_{2,HF}(s, r_{12}, r_{1'2})d\mathbf{r}_2 . \tag{15}$$

To this aim, replacing $\rho_{2,HF}(s, r_{12}, r_{1'2})$ in Eq. (15), the expression of $B_{Corr}(s)$ can naturally be recast in four contributions $B_{m,n}^{Corr}(s)$, reflecting orders $m$ and $n$ of correlation and exchange terms involved, respectively:

$$B_{1,1}^{Corr}(s) = \frac{2}{\rho}\int [-f(r_{12}) - f(r_{1'2})]\frac{1}{2}\rho^2 B_{HF}(r_{11'})d\mathbf{r}_2 , \tag{16}$$

$$B_{1,2}^{Corr}(s) = \frac{2}{\rho}\int [-f(r_{12}) - f(r_{1'2})]\frac{1}{2}\rho^2\left[-\frac{1}{2}B_{HF}(r_{12})B_{HF}(r_{1'2})\right]d\mathbf{r}_2 ,$$

$$B_{2,1}^{Corr}(s) = \frac{2}{\rho}\int [f(r_{12})f(r_{1'2})]\frac{1}{2}\rho^2 B_{HF}(r_{11'})d\mathbf{r}_2 \text{ and}$$

$$B_{2,2}^{Corr}(s) = \frac{2}{\rho}\int [f(r_{12})f(r_{1'2})]\frac{1}{2}\rho^2\left[-\frac{1}{2}B_{HF}(r_{12})B_{HF}(r_{1'2})\right]d\mathbf{r}_2 .$$

By construction of the correlation factor $f(r_{12})$, the very first contribution is zero, *i.e.* $B_{1,1}^{Corr}(s) = 0$, as pointed out in Ref. [7].

Regarding the remaining terms in Eqs. (16), the integrals involved can once more be approximated using the same ansatz as in Eq. (10). In addition, and as usual in similar problems [17,18], the main difficulty arises from integration of the term $r_{12}r_{1'2} = |\mathbf{r}_2 - \mathbf{r}_1||\mathbf{r}_2 - \mathbf{r}_1'|$, stemming from $f(r_{12})f(r_{1'2})$ in Eq. (15). Yet, this difficulty can be overcome thanks to prolate spheroidal coordinates, as defined e.g. in Ref. [19]. In particular, making use of $\mu = (|\boldsymbol{\sigma} + \mathbf{s}/2| + |\boldsymbol{\sigma} - \mathbf{s}/2|)/s$ and $\nu = (|\boldsymbol{\sigma} + \mathbf{s}/2| - |\boldsymbol{\sigma} - \mathbf{s}/2|)/s$, where $\boldsymbol{\sigma} = \mathbf{r}_2 - \mathbf{R}$, the integration of a given symmetric function $F(s, r_{12}, r_{1'2})$ over $\mathbf{r}_2$ amounts to calculate

$$\int F(s, r_{12}, r_{1'2})d\mathbf{r}_2 = \int_1^\infty d\mu \int_{-1}^1 \left(2\pi\frac{s^3}{8}(\mu^2 - \nu^2)F\left(s, \frac{s}{2}(\mu - \nu), \frac{s}{2}(\mu + \nu)\right)\right)d\nu , \tag{17}$$

which here meets an analytical solution. Skipping details of the calculation, the final result obtained for each component of the correlation contribution is

$$B_{1,1}^{Corr}(s) = 0, \tag{18}$$

$$B_{1,2}^{Corr}(s) = \frac{1225 r_s e^{-\frac{x^2}{10}}\left[8\sqrt{\frac{35}{\pi}}\left(8e^{\frac{7x^2}{320}} - 3\right)x - 3e^{\frac{7x^2}{320}}(7x^2 + 160)\text{erf}\left(\frac{1}{8}\sqrt{\frac{7}{5}}x\right)\right]}{6144 x\left(70 r_s + 3\sqrt[3]{2}\, 3^{2/3}\sqrt{35}\pi^{5/6}\right)}$$

$$B_{2,1}^{Corr}(s) =$$

$$\left\{1225 r_s^2 e^{-\frac{5x^2}{14}}\left(2e^{\frac{9x^2}{70}}\mathrm{erf}\left(\frac{3x}{\sqrt{70}}\right)\left\{2450\pi\mathrm{erfc}\left(\frac{3x}{\sqrt{70}}\right)e^{\frac{9x^2}{70}} + \sqrt{2}\left(3\sqrt{35\pi}x - 140\right)\left(9x^2 + 35\right)\right\} - 3x\left\{\sqrt{\frac{70}{\pi}}\left(9\pi x^2 + 35\pi - 560\right)e^{\frac{9x^2}{70}} + 140\left(4\sqrt{\frac{35}{\pi}} - 3x\right)\right\}\right)\right\} \bigg/ \left\{1944 x\left(70 r_s + 3\sqrt[3]{2}\, 3^{2/3}\sqrt{35}\pi^{5/6}\right)^2\right\}$$

and

$$B_{2,2}^{Corr}(s) =$$

$$-\left\{49 r_s^2 e^{-\frac{5x^2}{14}}\left(6\mathrm{erf}\left(\frac{1}{2}\sqrt{\frac{5}{7}}x\right)\left\{588\pi\mathrm{erfc}\left(\frac{1}{2}\sqrt{\frac{5}{7}}x\right)e^{\frac{5x^2}{28}} + 3\left(3\sqrt{35\pi}x - 140\right)\left(5x^2 + 14\right)\right\}e^{\frac{5x^2}{28}} + x\left\{e^{\frac{5x^2}{28}}\sqrt{\frac{35}{\pi}}\left(2800 - 9\pi\left(5x^2 + 14\right)\right) - 420\left(4\sqrt{\frac{35}{\pi}} - 3x\right)\right\}\right)\right\} \bigg/ \left\{600 x\left(70 r_s + 3\sqrt[3]{2}\, 3^{2/3}\sqrt{35}\pi^{5/6}\right)^2\right\}$$

wherein $x = k_F s$. Finally, the normalization factor $\eta$ is obtained from the limit $B(0) = 1$ in Eq. (14), leading to

$$\eta = \tag{19}$$

$$\left\{259200\sqrt{\pi}\left(70 r_s + 3\sqrt[3]{2}\, 3^{2/3}\sqrt{35}\pi^{5/6}\right)^2\right\} \bigg/ \left\{7\left(1176\sqrt{35}\left(3125\sqrt{2} - 972\right)\pi\, r_s^2 + 181440000\sqrt{\pi}\, r_s^2 + 490\sqrt{35}\left(40000\sqrt{2} - 59669\right)r_s^2 + 15552000\sqrt[3]{2}\, 3^{2/3}\sqrt{35}\,\pi^{4/3} r_s + 12403125\sqrt[3]{2}\, 3^{2/3}\pi^{5/6} r_s + 349992000\, 2^{2/3}\sqrt[3]{3}\,\pi^{13/6}\right)\right\}$$

The density-matrix underlying the RC functional is now entirely determined by Eqs. (13) - (19). It shall hereafter be referred to as $B_{RC}(s)$. It can be verified that $B_{RC}(s)$ yields the kinetic energy as given in Eq. (11).

## III.2. Density-matrix from parameterized momentum distributions

Other closed-form expressions of accurate density-matrices may suitably be obtained from parameterized momentum distributions, provided the latter is simple enough to allow for analytical inversion. In this respect, a well accepted example of parameterized momentum distribution is that provided by Barbiellini and Bansil [20], that is,

$$n_{BB}(p) = \tag{20}$$

$$\begin{cases} a_1\left(1 - a_2\, y^2\right), & \text{if } y < 1 \\ a_3 \exp(-a_4\,(y-1)) + T/y^8 & \text{otherwise,} \end{cases}$$

where $y = p/k_F$ and $a_1$ - $a_4$ are fitting parameters that find simple $r_s$-dependent expressions. Following the method used in Ref. [20], the last coefficient can be taken as $T = \frac{4}{9}\left(\left(\frac{4}{9\pi}\right)^{1/3}\frac{r_s}{\pi}\right)^2 g(0)$, where $g(0)$ is the pair correlation function at $r_{12} = 0$, *i.e.* the "on-top pair density", which may for instance be taken as in Ref. [21].

Advantageously, Eq. (20) transforms back to position space, namely via the relation

$$B(s) = \int n(p)\, \frac{4\pi p \sin(ps)}{s}\, dp, \tag{21}$$

leading to

$$B_{BB}(s) = \quad (22)$$

$$\frac{1}{160 x^5 r_s^3} \pi^2 \{T\, x^4 (-(\pi - 2\,\mathrm{Si}(x))x^6 + 2(x^4 - 2x^2 + 24)x\cos(x) + 2(x^4 - 6x^2 + 120)\sin(x))$$

$$+ \frac{1}{(x^2 + a_4^2)^2} 1440 a_3 x^4 (\sin(x)((a_4 - 1)x^2 + a_4^2(a_4 + 1)) + x\cos(x)(x^2 + a_4(a_4 + 2))) +$$

$$1440 a_1 (x\cos(x)((a_2 - 1)x^2 - 6a_2) + \sin(x)(x^2 - 3a_2(x^2 - 2)))\}$$

where Si($x$) stands for the sine integral function [19]. The result of Eq. (22), i.e. $B_{BB}(s)$, shall be compared with $B_{RC}(s)$ in the next sections.

Finally, a more elaborated parameterization of the momentum distribution has been proposed in a work of Gori-Giorgi and Ziesche [22]. Yet, while the said parameterization is believed to be one of the most accurate to date, it hardly permits an analytical inversion. A numerical transform can yet be performed. It shall furthermore be useful for comparing the momentum distributions obtained *in fine*.

## IV. Comparison of correlated density-matrices

### IV.1. Asymptotic expansions

A variety of properties of both the momentum distribution and the density-matrix are discussed in Ref. [22]. In particular, the small- and large-$s$ expansion of the exact density-matrix are given as

$$B_{Exact}(s) = 1 - \frac{1}{3!}\langle p^2 \rangle s^2 + \frac{1}{5!}\langle p^4 \rangle s^4 - \frac{1}{5!}\pi^4 T \rho (k_F s)^5 \ldots \quad (23)$$

and

$$B_{Exact}(s) = -3 z_F \frac{\cos(k_F s)}{(k_F s)^2} + O\left(\frac{1}{s^3}\right), \quad (24)$$

where $T = \frac{4}{9}\left(\left(\frac{4}{9\pi}\right)^{1/3} \frac{r_s}{\pi}\right)^2 g(0)$ as exemplified earlier, $z_F$ is the $r_s$-dependent Fermi gap and $\langle p^n \rangle$ denotes the $n$th momentum moment.

It has been verified that both $B_{RC}(s)$ and $B_{BB}(s)$ have similar formal series developments as in Eqs. (23) and (24), from which respective momentum moments, on-top pair densities or Fermi gaps can be derived. It is therefore interesting to compare such quantities.

First, concerning the small-$s$ expansion: second-order moments $\langle p^2 \rangle$ (or twice the kinetic energy) as derived from $B_{RC}(s)$ and $B_{BB}(s)$ are depicted in Fig. 1. The moments obtained from $B_{BB}(s)$ are obviously in excellent agreement with Perdew and Wang's [9], at least for $r_s$ = 2 to 5, which is the assumed range of validity of the parameterization of Ref. [20]. The second momentum moments obtained from $B_{RC}(s)$ generally lie intermediary between those derived from $B_{HF}(s)$ and $B_{BB}(s)$. In this regards, it was already noted in Ref. [7] that the RC kinetic energy underestimates Perdew and Wang's. However, the agreement with Perdew-Wang's results substantially improves at lower $r_s$ ($r_s < 2$).

The fourth momentum moments $\langle p^4 \rangle$ derived from $B_{RC}(s)$ and $B_{BB}(s)$ show a much better agreement with each other, and they further both clearly depart from the HF results, as shown in Fig. 2. No additional reference data is considered for comparison here.

The utmost departure from the HF theory occurs in fact for the next term in the series expansion of $B_X(s)$. Indeed, the $s^5$-term in the expansion of $B_{HF}(s)$ is zero, in contrast to $B_{RC}(s)$, $B_{BB}(s)$, or the exact expansion of Eq. (23). Thus, the $s^5$-term is in its entirety linked to Coulomb correlation.

In Fig. 3, the on-top pair densities resulting from the small-$s$ expansion of $B_X(s)$ are compared with the results reported in the work of Gori-Giorgi and Perdew [23]. As can be seen, the resulting on-top densities are much the same except at large $r_s$, whereas the agreement with Gori-Giorgi and Perdew's results [23] is less satisfactory. This does matter inasmuch as $g(0)$ determines the large-$p$ behavior of the momentum distribution, see e.g. [23]. In this respect, in spite of incorrect asymptotic behaviors, the RC model (Fig. 3) involves a fairly good description of the on-top pair density and improves over the parameterized model of Ref. [20], at least in the range considered in Fig. 3. This should in turn result in a reasonable large-$p$ behavior for the corresponding momentum distribution.

Second, by construction, $B_{RC}(s)$ has a large-$s$ asymptote formally comparable with that of $B_{Exact}(s)$. Indeed, at large $s$, $B_{RC}(s)$ is dominated by $B_{RC}(s) \approx \eta B_{HF}(s)$, see Eqs. (16) and (18), such that $\lim B_{RC}(s)|_{s \to \infty} = -3\eta \frac{\cos(k_F s)}{(k_F s)^2}$, in agreement with Eq. (24). Accordingly, in the RC approach, the $r_s$-dependent normalization factor $\eta$ plays the role of the Fermi break function. Thus, a direct link is established between the normalization of the underlying correlated wavefunction and the Fermi break.

In Fig. 4, the Fermi break functions emerging from the large-$s$ expansions of $B_{BB}(s)$ and $B_{RC}(s)$ are compared with $z_F$ as parameterized in Ref. [23]. In spite of the correct limit at $r_s = 0$ and the global decreasing trends, a substantial disagreement occurs. In particular, the RC model markedly overestimates the results of Ref. [23], which should accordingly impair the corresponding momentum distribution near the Fermi momentum.

In this respect, note that the gaussian tail of the correlation factor $f(r_{12})$ used in the RC approach prevents any other terms under the integral sign of Eqs. (15) or (16) from contributing to the exact limit of Eq. (24). Thus, the correlation factor tail necessarily impairs the Fermi break in the RC model.

### IV.2. Comparisons of density-matrices

Beyond asymptotic expansions, $B_{RC}(s)$ as determined by Eqs. (13) - (19) can be directly compared with the expression of $B_{BB}(s)$ as defined in Eq. (22). As noted earlier, it is hardly possible to safely estimate the position-space counterpart of the parameterized momentum distribution of Gori-Giorgi and Ziesche [22], even numerically, as correlation effects extend far in momentum space. A numerical inversion has nevertheless been considered here; the result, $B_{GZ}(s)$, shall thus serve as a reference.

Yet, it is actually preferable to compare difference functions like $\Delta B_X(s) = B_X(s) - B_{HF}(s)$ to emphasize the correlation effects, which else are hardly visible. Such differences are shown in Figs. 5A and B, respectively corresponding to the cases $r_s = 2$ and 5. Note that $B_{mRC}(s)$ refers to a simple variant to the RC model that shall be discussed later.

The functions $\Delta B_X(s)$ are all zero at the origin, owing to the normalization condition. In other aspects, they resemble a usual waveform, which oscillates to zero as $s$ increases. Most notably, all curves qualitatively agree. Comparing $\Delta B_{RC}(s)$ and $\Delta B_{BB}(s)$ with respect to $\Delta B_{GZ}(s)$, it turns that $\Delta B_{BB}(s)$ shows a better phase agreement at small $s$. However, $\Delta B_{RC}(s)$ provides amplitudes closer to $\Delta B_{GZ}(s)$, especially for $r_s = 5$. Incidentally, the quantitative agreement of $\Delta B_{RC}(s)$ and $\Delta B_{BB}(s)$ improves as $s$ increases. Despite the quantitative misalignment at small $s$, it is worth reminding that the curvatures of the depicted functions near the origin should nevertheless be much similar, in view of the results commented in reference to Fig. 1.

Altogether, the RC matrix performs nicely, provided that $\Delta B_{GZ}(s)$ can safely serve as the reference. This is however subjected to numerical errors that necessarily occur when inverting $n_{GZ}(p)$, Ref. [22], especially as correlation effects extend far in momentum space. Incidentally, considering the parameterization proposed in Ref. [24] for $n_{BB}(p)$ instead of that of Ref. [20] does not substantially change the above conclusions.

### IV.3. Momentum distributions

It was so far not possible to obtain a closed-form result for the 3D Fourier transform of $B_{RC}(s)$. However, the component $B_{Corr}(s)$, see Eqs. (18), is well localized in position space, owing to the gaussian ansatz. Thus, a numerical integration allows its momentum-space counterpart to be evaluated safely.

In Figs. 6A and B, the resulting momentum distribution $n_{RC}(p)$ is compared with $n_{BB}(p)$, Eq. (20), and with the parameterized model $n_{GZ}(p)$ of Ref. [22]. To this aim, a Fortran subroutine kindly made available on-line was used [25]. Again, only the cases $r_s = 2$ and 5 are illustrated.

Although the obtained $n_{RC}(p)$ is roughly realistic, it is immediately visible that $n_{RC}(p) > 1$ for some values of $p < p_F$. Thus, eigenvalues of the density-matrix are not systematically in the range 0 – 1, whereby its $N$-representability condition is manifestly infringed [26]. This is however not surprising in view of the approximations inherent to the model.

Clearly, both the truncation of the one-matrix expansion, Eq. (7) and the gaussian ansatz used for $\rho_{2,HF}(s, r_{12}, r_{1'2})$ in Eq. (15) deprive the model of its $N$-representability.

### IV.4. Simple correction to the RC model

In particular, the gaussian ansatz to the HF ingredients is accurate at small $r_{12}$ only, not at large separation distance. This indirectly impairs the model at large $s = r_{11'}$, after integration over $\mathbf{r}_2$ in Eqs. (15) - (16), which in turn unfavorably impacts the momentum distribution, notably at small $p$. Also, introducing a gaussian component in the position space amounts to introducing a gaussian component also in the momentum space, whence the bell-shape of $n_{RC}(p)$ for $p < p_F$, see Figs. 6A and B.

Therefore, one may want to refine the initial ansatz, i.e., $B_{HF}(s) \approx e^{-\beta_x^2 s^2}$. For instance, considering instead $B_{HF}(s) \approx e^{-\beta_x^2 s^2}(1 - \beta_x^2 s^2 / 7)$ matches the exact HF result up to fourth order. Then, repeating the calculation described earlier with respect to the initial RC scheme changes the final expression obtained for the correlation part $B_{Corr}(s)$ and in turn the kinetic energy. For the sake of comparison, the same value as in the initial RC approach was kept for $q = 3(3\pi^2)^{1/3}/\sqrt{35}$, the sole parameter involved in the RC approach. This variant is referred to as "mRC". The final expression of $B_{mRC}(s)$ is much more cumbersome than Eqs. (15) – (19) and is therefore not given here.

The difference $\Delta B_{mRC}(s)$ is represented in Figs. 5A and B. As can be seen, it slightly improves the phase and the amplitude of $\Delta B_{RC}(s)$, with respect to either $\Delta B_{BB}(s)$ or $\Delta B_{GZ}(s)$, especially at small $s$. The improvement brought by the mRC variant is in fact most visible on the momentum distribution, Figs. 6A and B, where the condition $n_{mRC}(p) \leq 1$ is now mostly satisfied.

Such results strongly support the assumption that the initial gaussian ansatz is by and large responsible for the non-representability of the RC model, at least for the electron densities considered here. Obviously, since the terms neglected in the expansion of Eq. (5) involve increasing numbers of electrons in a same region, they must play an increasing role when $r_s \to 0$. The mRC variant does otherwise not substantially improve on the initial RC near the Fermi momentum or above.

Interestingly, in the mRC approach, the correlation kinetic energy becomes

$$t_{c,mRC}(r_s) = \frac{1 + 0.0231261/r_s}{11.7858 + 14.7411 r_s + 4.80054 r_s^2} \tag{25}$$

and is depicted in Fig. 7, where it is furthermore compared with the results of Perdew and Wang, and the initial RC model. The results of Ref. [20] are reported too, for completeness; it is however reminded that the parameterization it uses is not meant to be valid for $r_s < 2$.

The new correlation kinetic energy is very similar to that of the initial RC approach, except at small $r_s$, owing to the $1/r_s$ term in the numerator. Thus, the high-density limit of $t_{c,mRC}(r_s)$ is still incorrect in comparison with the theoretical result $t_c|_{r_s \to 0} = \frac{(\ln 2 - 1)(\ln r_s + 1)}{\pi^2}$, see e.g. Ref. [27]. However, it strikingly improves on the RC model in the range $r_s = 0.01 - 1$. In fact, in this range, the mRC model renders 99.7% of the Perdew-Wang result on average, vs. 89.2% for the initial RC model.

This suggests that the improvement provided by the modified mRC approach is not fortuitous. However, as it stands, the expression of $t_{c,mRC}(r_s)$ given in Eq. (25) has still several drawbacks. First, the offered improvement restricts to $r_s = 0.01 - 1$. Second, the parameter $q = 3(3\pi^2)^{1/3}/\sqrt{35}$ was straightforwardly imported from the initial RC model instead of being optimized. Third, the $1/r_s$ dependence at numerator results in an improper behavior of the correlation energy at small $r_s$ (by inversion). Thus, a more elaborate improvement to the initial RC model is still needed but this is outside the scope of the present work.

In defense of the initial RC model, one may argue that the fraction of electrons infringing the *N*-representability condition (see Figs. 6A and B) is not that significant (typically < 1%), and the impact on the kinetic energy density is even less (< 0.2 %). Indeed, inspection of the kinetic energy distributions ($T_X(p) \propto p^4 n_X(p)$) for $r_s = 2 - 5$ shows that the distributions $T_{RC}(p)$ and $T_{BB}(p)$ are not distinguishable at low momenta ($p < p_F$), even near the Fermi edge, see Fig. 8. Note that, in Fig. 8, also the HF kinetic energy distribution has been included, for comparison.

In addition, although not visible at the scale of Figs. 6A and B, the RC model provides much more realistic results at larger momenta ($p > p_F$), consistently with previous remarks in reference to Fig. 3, as to the $s^5$ dependency in the expansion of corresponding position-space functions. In particular, for some sufficiently large values of $p > p_F$, the RC model competes with the parameterized momentum distribution of Ref. [20], considering the model $n_{GZ}(p)$ as the exact momentum-density [22] (not visible at the scale of Figs. 6A and B). This is consistent with the comments to Figs. 5A and B.

# Final remarks

The density-matrix model underlying the RC approach was derived in closed form. It was shown to provide a reasonably good agreement with counterparts obtained from parameterized momentum distributions. On the face of it, the main disagreement is rather quantitative than qualitative. For instance, the expansion of the RC matrix at large-$s$ values is qualitatively but not quantitatively correct. As a result, the Fermi break function is substantially in error. On the contrary, momentum moments derived from the small-$s$ expansion of the RC matrix are quite accurate.

Turning to momentum space, the RC model is *prima faciae* more questionable. In particular, the *N*-representability condition appears to be infringed for the smallest values of $p < p_F$. While one may first put into question the truncation in the expansion of the density-matrix model, this drawback is most likely a consequence of the gaussian ansatz used for the Hartree-Fock exchange ingredient of the RC model. Results obtained from a simple modification ("mRC") of the the gaussian ansatz in the RC model support this assumption.

Nevertheless, the fraction of infringing electrons in the RC model is small and pertains to slowest electrons only, with no meaningful impact on the kinetic energy or higher momentum-moments properties. In addition, the momentum distribution itself is nicely rendered at larger momenta.

Apart from limitations as to its *N*-Representability, the RC model remains based on a properly normalized density-matrix, which itself derives from a correlated wavefunction, subject to approximations. From this point of view at least, it benefits from a satisfactory physical background. Now, its drawbacks as to the description of the UEG may turn advantageous in LSDA applications. In this respect, long-range correlation effects in the UEG are certainly hardly transposable to finite systems, at least within LSDA. In the RC model, long-range correlation is damped by gaussian factors; which notably impact the conjugated effects of exchange and correlation, see e.g. $B_{m,1}^{Corr}(s)$ *vs.* $B_{m,2}^{Corr}(s)$ in Eq. (16). Hence, the local energy expression obtained discards a substantial part of what occurs in a UEG but not in finite systems. This may at least partly explain improvements of the RC model over the standard LSDA.

Hence, keeping in mind that the RC model is otherwise fully analytic, simple, wavefunction-based and parameter-free, it certainly provides an interesting basis for local and gradient-corrected approximations to the correlated properties of atoms and molecules, as illustrated in Refs. [7, 10 - 12].

# Figure captions

Fig. 1: Second momentum moments of the electron gas, as obtained from the expansions of $B_{RC}(s)$ (full line), $B_{BB}(s)$ (dashed), $B_{HF}(s)$ (dotted), and 'exact' results of Perdew and Wang (grey), Ref. [9].

Fig. 2: Fourth momentum moments of the electron gas obtained from $B_{RC}(s)$ (full line), $B_{BB}(s)$ (dashed), and $B_{HF}(s)$ (dotted).

Fig.3: On-top pair density of the electron gas obtained from $B_{RC}(s)$ (full line), $B_{BB}(s)$ (dashed), using Eq. (23), and Ref. [23] (dotted).

Fig. 4: Fermi-break functions of the electron gas obtained from $B_{RC}(s)$ (full line), $B_{BB}(s)$ (dashed), and from Ref. [23] (dotted).

Fig. 5A - B: Representation of the difference $\Delta B_X(s) = B_X(s) - B_{HF}(s)$, for $r_s = 2$ (Fig. 5A) and 5 (Fig. 5B). Full line: $X = RC$; grey: $X = mRC$; dashed: $X = BB$; dotted: $X = GZ$, obtained by numerical transform of $n_{GZ}(p)$ as provided in Ref. [22].

Figs.6A – 6B: Momentum distributions for $r_s = 2$ (Fig. 6A) and 5 (Fig. 6B). Full line: $n_{RC}(p)$; grey: $n_{mRC}(p)$; dashed: $n_{BB}(p)$; dotted: $n_{GZ}(p)$, calculated as in Ref. [22].

Fig. 7: Representation of the correlation kinetic energy *vs.* $r_s$. Full line: from the $n_{RC}(p)$ model. Dashed: from $n_{BB}(p)$, Ref. [20]. Dotted: Perdew and Wang, Ref. [9]. Grey: $n_{mRC}(p)$.

Fig. 8: Kinetic energy distributions $T_X(p) \propto p^4 n_X(p)$ for $r_s = 2$. Full line: $T_{RC}(p)$, dashed: $T_{BB}(p)$. Dotted: $T_{GZ}(p)$, calculated as in Ref. [22]. Grey: $T_{HF}(p)$.

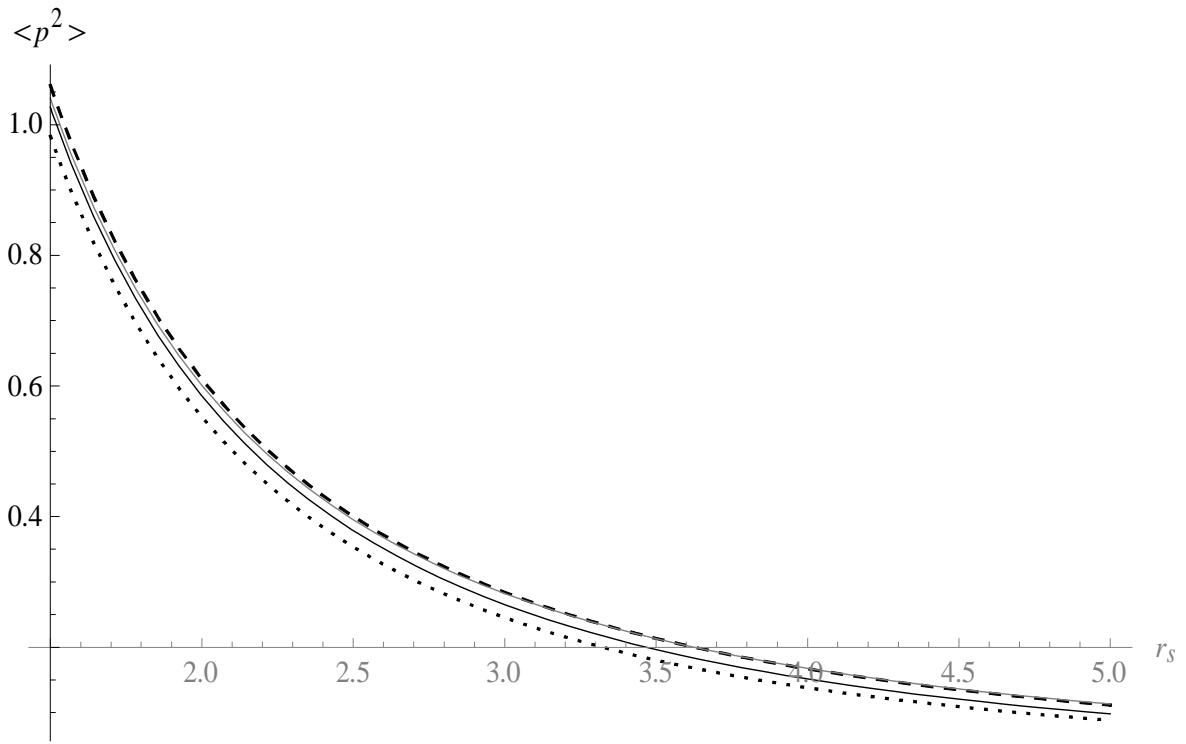

**FIG. 1.**

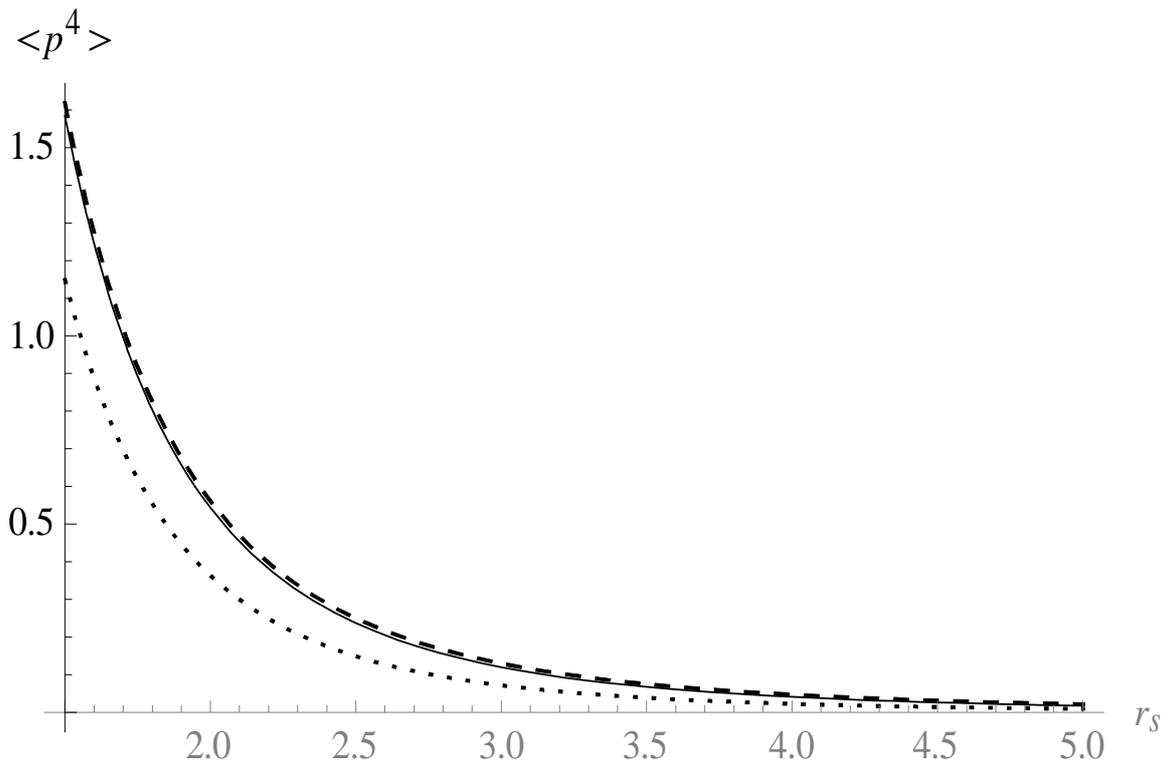

**FIG. 2**

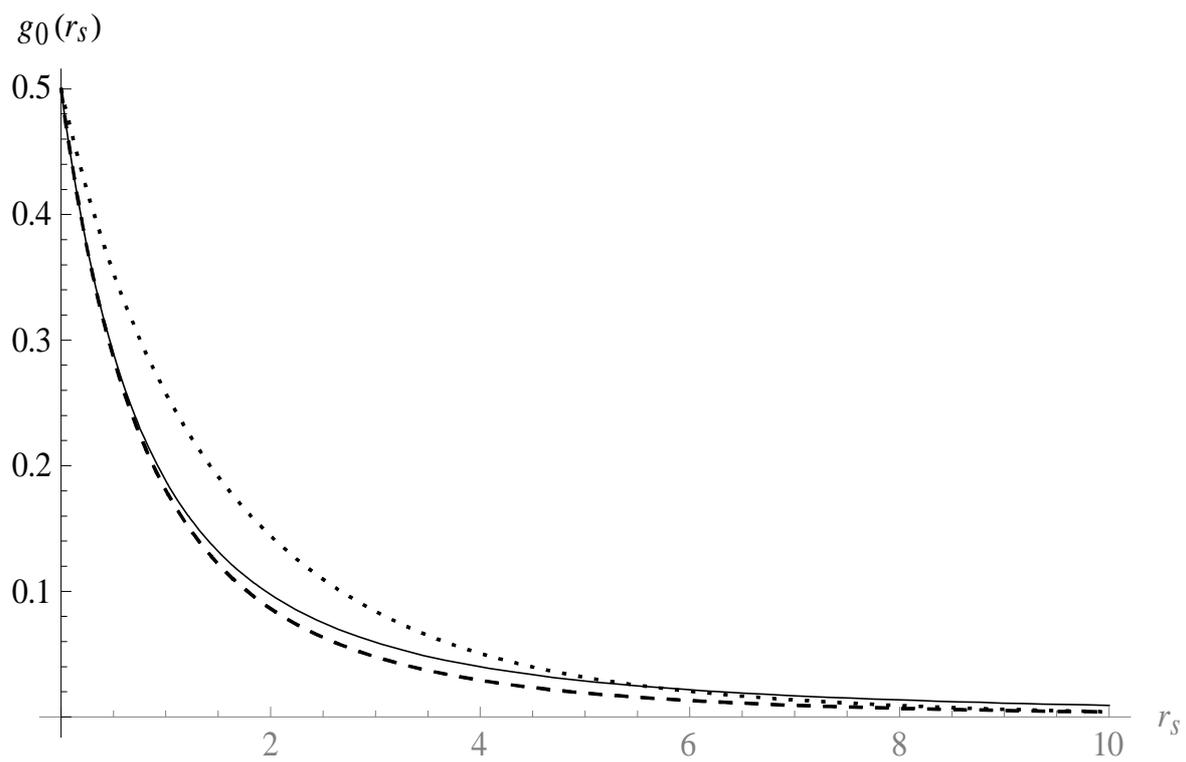

**FIG. 3**

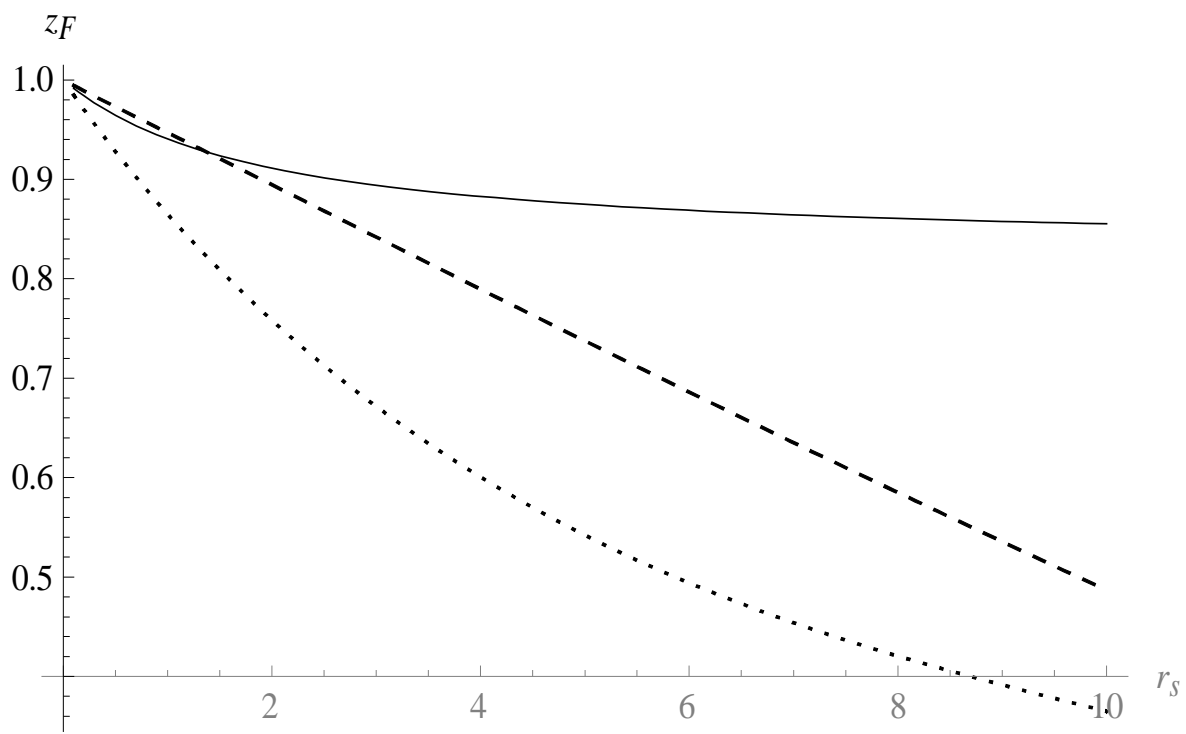

**FIG. 4**

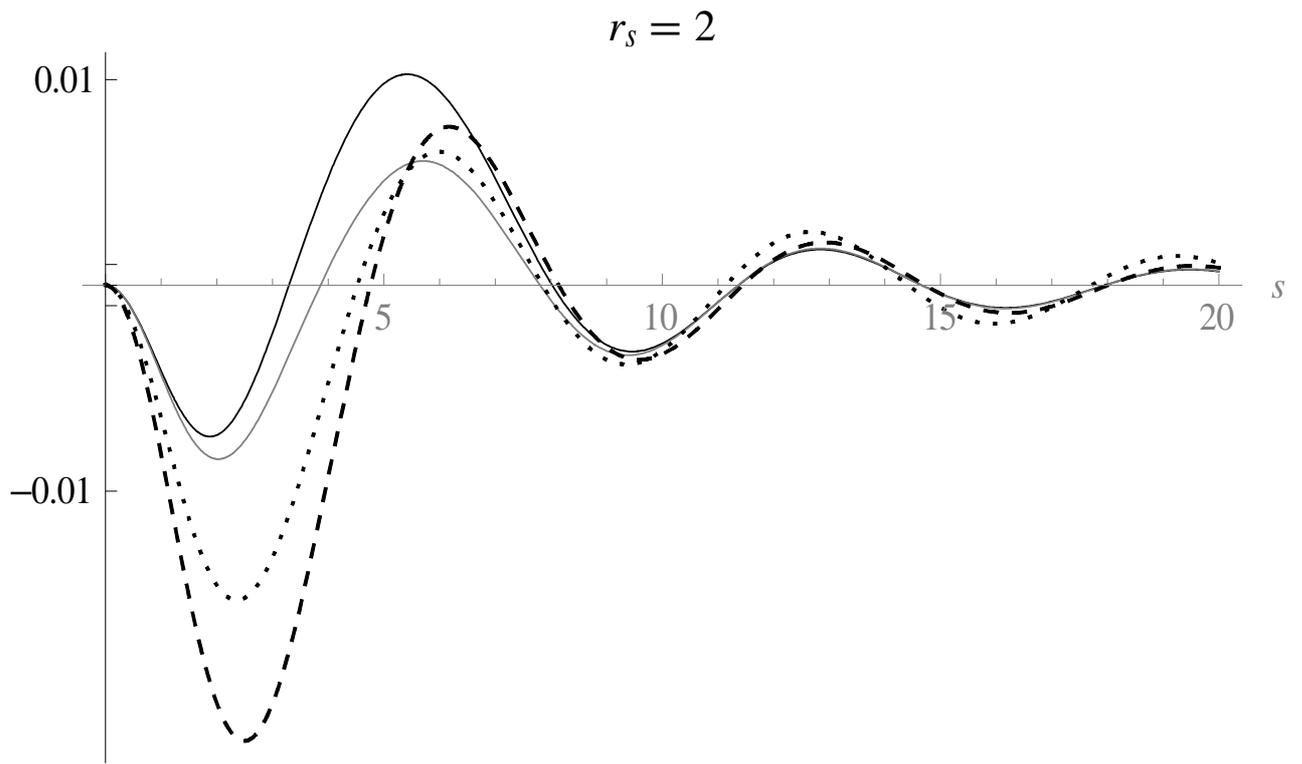

**FIG. 5A**

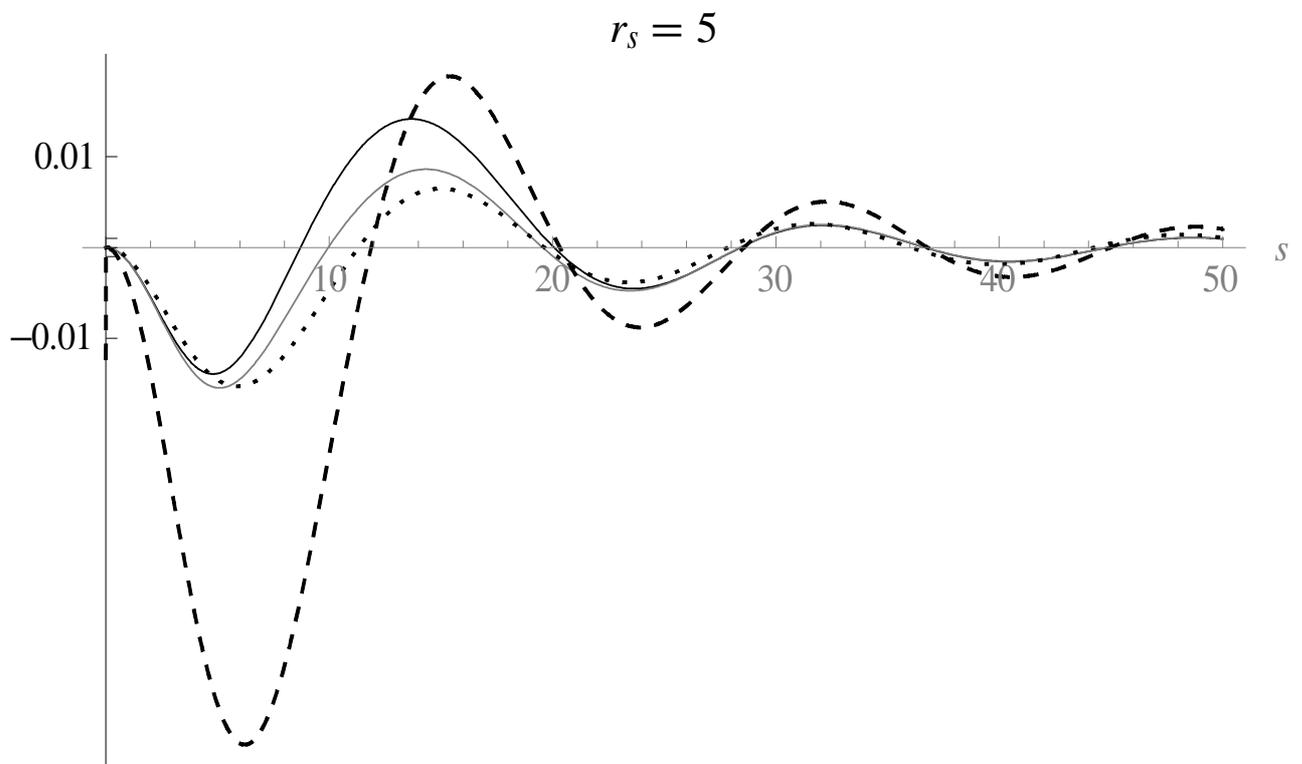

**FIG. 5B**

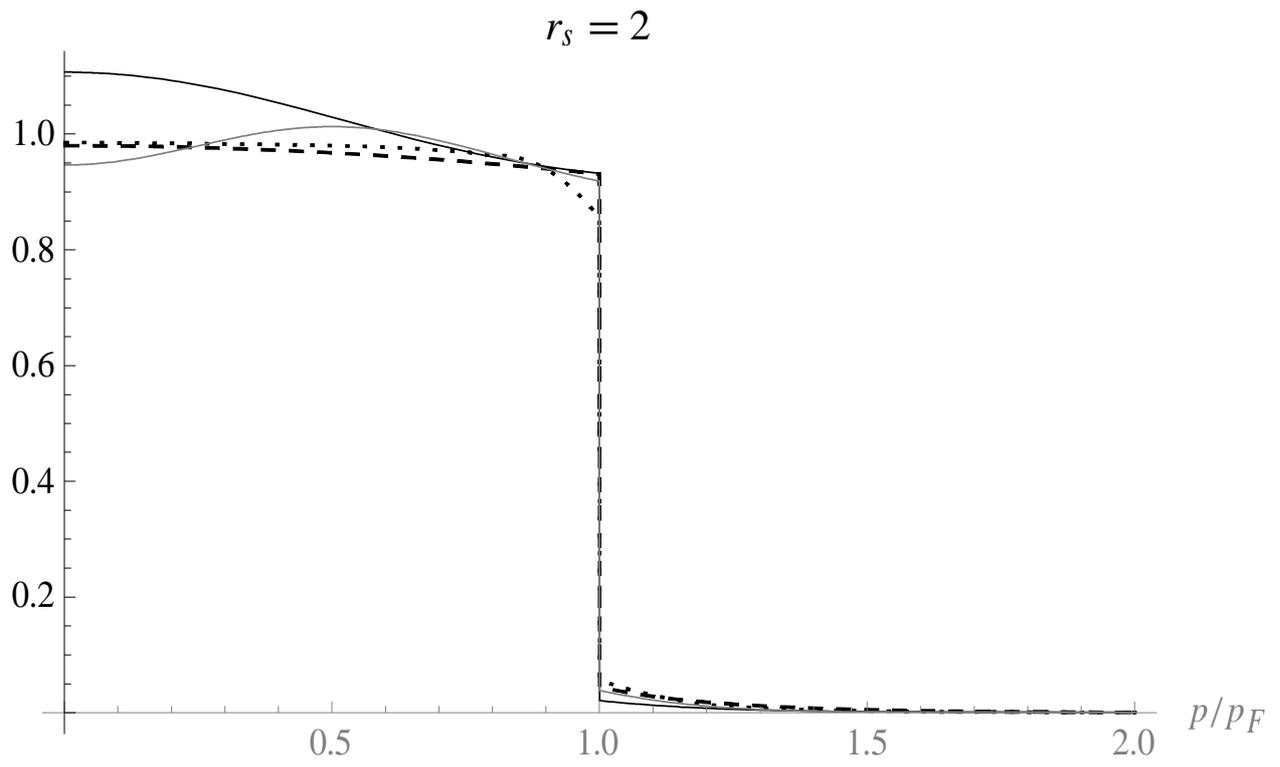
**FIG. 6A**

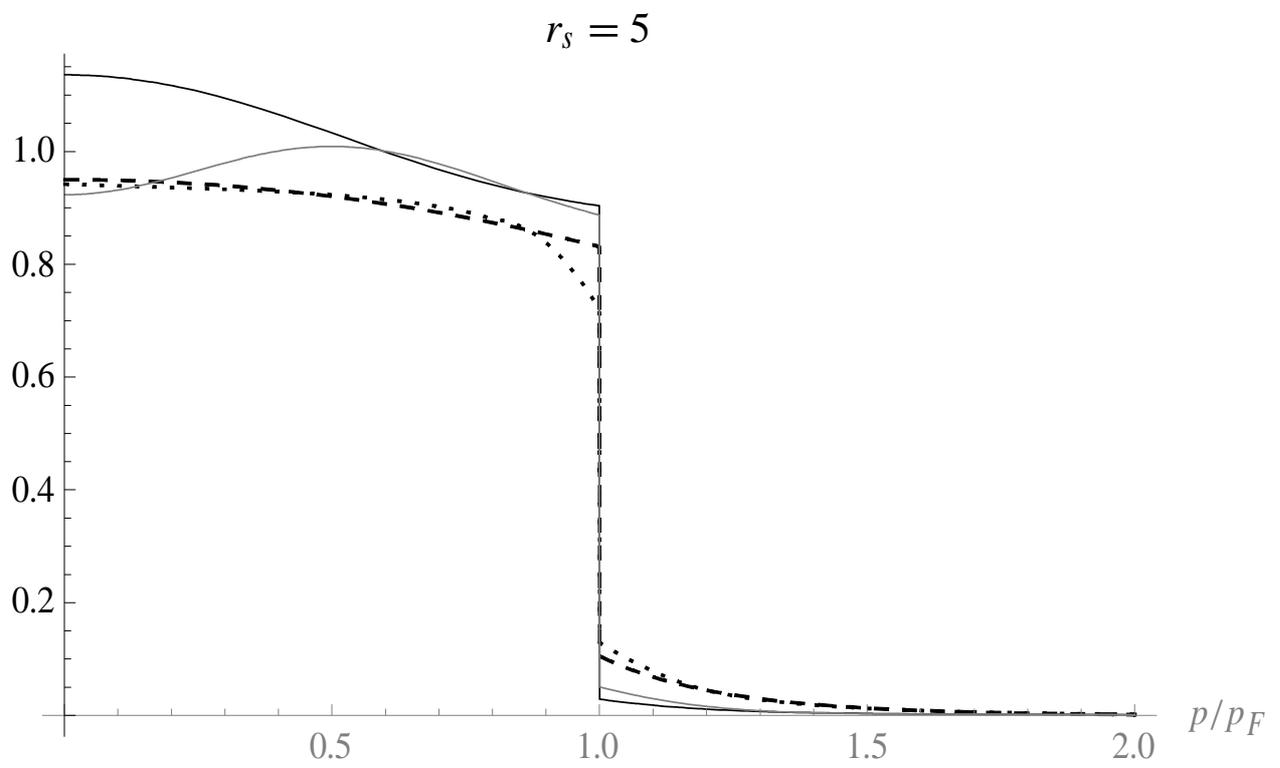
**FIG. 6B**

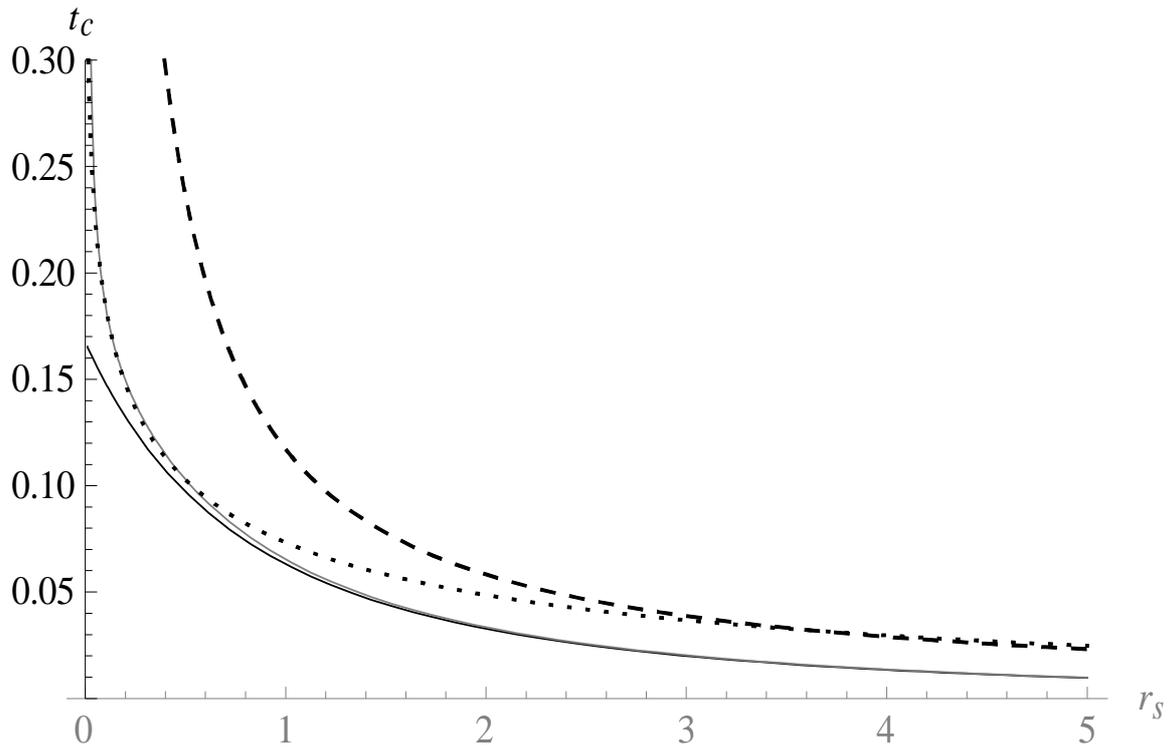

**FIG. 7**

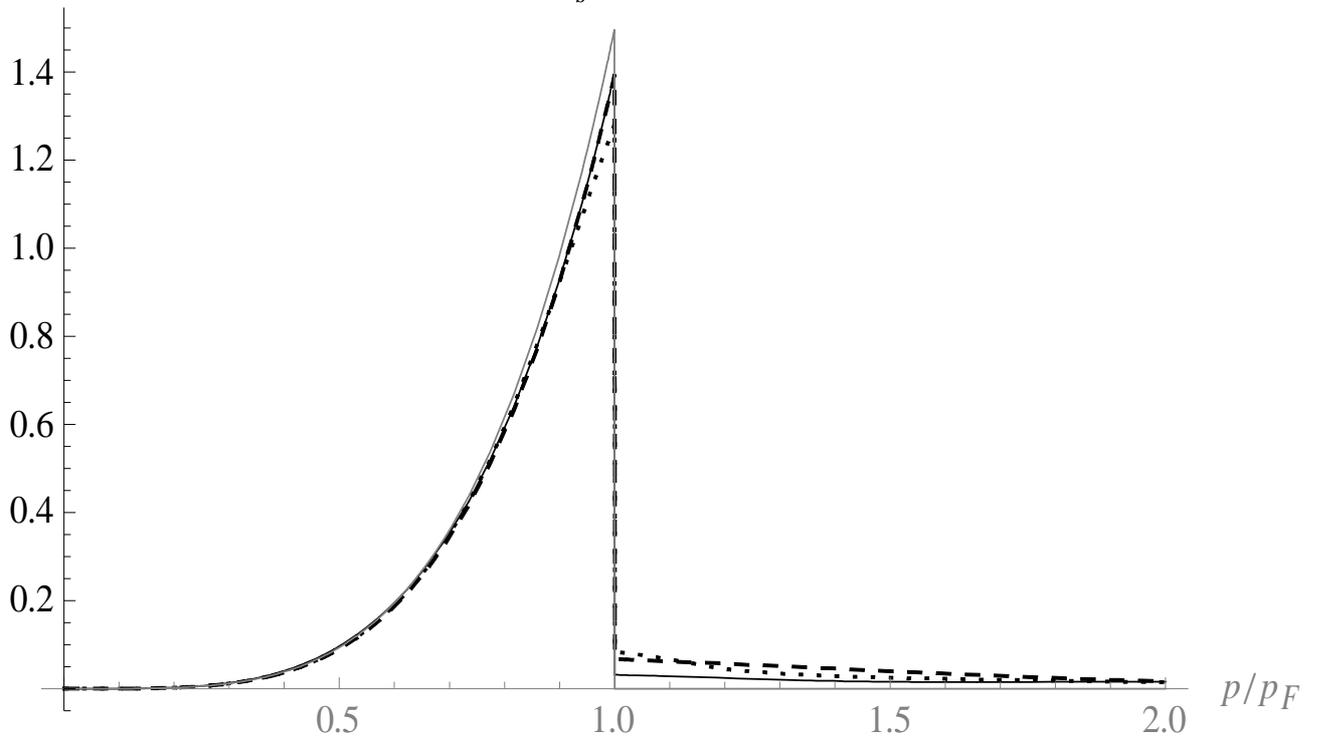

**FIG. 8**